\documentclass[twocolumn,showpacs,preprintnumbers,amsmath,amssymb]{revtex4}
\usepackage{graphicx}% Include figure files
\usepackage{dcolumn}% Align table columns on decimal point
\usepackage{bm}% bold math
 
\begin{document}
 
\preprint{APS/123-QED}
 
\title{Anomalous Fermi Liquid Behavior of Overdoped High-$T_c$ Superconductors}
 
\author{H. Castro}
\author{G. Deutscher}
\email{guyde@post.tau.ac.il}
\affiliation{School of Physics and Astronomy, Tel Aviv University}
 
\homepage{http://www.tau.ac.il/~supercon}
 
\date{\today}
 
\begin{abstract}
According to a generic temperature vs. carrier-doping ($T-p$) phase diagram of high-temperature superconductors it has been proposed that as doping increases to the overdoped region they approach gradually a conventional (canonical) Fermi Liquid. However, Hall effect measurements in several systems reported by different authors show a still strong \emph{T}-dependence in overdoped samples. We report here electrical transport measurements of $Y_{1-x}Ca_{x}Ba_{2}Cu_{3}O_{7-\delta}$ thin films presenting a temperature dependence of the Hall constant, $R_H$, which does not present a gradual transition towards the \emph{T}-independent behavior of a canonical Fermi Liquid. Instead, the \emph{T}-dependence passes by a minimum near optimal doping and then increases again in the overdoped region. We discuss the theoretical predictions from two representative Fermi Liquid models and show that they can not give a satisfactory explanation to our data. We conclude that this region of the phase diagram in YBCO, as in most HTSC, is not a canonical Fermi Liquid, therefore we call it Anomalous Fermi Liquid. 
\end{abstract}
 
%Use showkeys class option if keyword
\maketitle
 
\section{Introduction}
 
\subsection{Phase Diagram}
High temperature Superconductors (HTSC) are known to undergo fundamental changes in some of their properties as the carrier density (doping, \emph{p}) is changed. At low doping we have an insulating anti-ferromagnetic phase (AFM), possibly a Mott-Hubbard insulator, which gradually disappears upon increase of doping leaving place to an anomalous metallic phase presenting superconductivity above a certain doping level. Figure 1 presents such a generic phase diagram of HTSC for hole doping \cite{Battloggb, Orenstein}. Apart from the intensive research on the superconductive (SC) region initiated after the discovery of HTSC in 1986, the surrounding normal phases began to receive special attention during the last decade. The existence of a pseudogap (PG) below the temperature $T^*(p)$, is today a well established fact in the underdoped regime $p < p_{op}$ ($p_{op} \approx 0.15$), confirmed by different experimental techniques, like NMR, ARPES, specific heat measurements, tunnelling, etc. as reviewed in \cite{Timusk, Loram}. The 'strange metal', or Marginal Fermi Liquid (MFL) phase has been also well characterized experimentally, and given a detailed theoretical explanation \cite{Varma}. However, the Canonical Fermi Liquid (FL) phase, as defined in \cite{Levin}, supposed to exist below a temperature $T_f(p)$ \cite{Battloggb, Orenstein, Tsuei, Kaminski, Friedel} in the overdoped region ($p > p_{op}$), still rests on speculative grounds with contradictory experimental data and no clear theoretical understanding. We want to provide additional experimental data and discussion in order to shed more light on the nature of this region. Our main conclusion is that this phase is not a canonical FL. 

\begin{figure}
\includegraphics[width=0.5\textwidth]{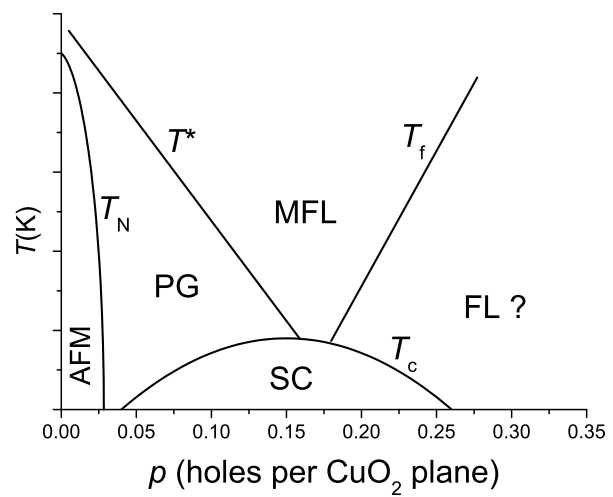}% Here is how to import EPS art
\caption{\label{fig:epsart} General $T-p$ phase diagram for hole doping in HTSC. Phases: Anti-ferromagnetic (AFM), pseudogap (PG), Marginal Fermi Liquid (MFL), superconductor (SC), Fermi Liquid (FL). The question mark in the FL phase indicates it is not confirmed.}
\end{figure}

\subsection{\label{sec:level2}Resistivity}
The temperature $T^{*}$ can be determined from resistivity  vs. temperature measurements as the point where the resistivity, $\rho$, departs (decreases) from a linear behavior in underdoped samples \cite{Battlogg}. The drop of $\rho$ is understood to be a consequence of the opening of a PG in the interaction at the origin of carriers scattering, for instance spin fluctuation \cite{Bucher,Ito}. In a similar way $T_f$ can be obtained from resistivity measurements as the departure from linearity (increase) in overdoped samples \cite{Kaminski}. The existence of a certain anomaly along the line $T_f$ has been theoretically suggested by Friedel and Kohmoto\cite{Friedel}, and a new phase below the same line was demonstrated by ARPES measurements \cite{Kaminski}. The upturn in $\rho$ vs. \emph{T} has been fitted to $\rho = a+bT^2$ \cite{Pines}, which would imply the emergence of FL behavior assuming a dominant fermion-fermion scattering \cite{Levin}. However, this conclusion is not quite clear since the upturn occurs at temperatures much higher than in usual metals. Furthermore, deviations from the above quadratic law have been observed, which contradict the FL assumption. For instance, Proust et al. \cite{Proust} have reported a behavior $\rho = \rho_{o}+\alpha T+\beta T^{2}$ for Tl-2201 with a substantial linear term ($\alpha T > \beta T^{2}$). Naqib et al. \cite{Naqib} reported also a dependence $\rho = \rho_{o}+a T^{m}$ for their measured samples $Y_{1-x}Ca_{x}Ba_{2}(Cu_{1-y}Zn_{y})_{3}O_{7-\delta}$, as well as in samples measured by other authors: $La_{2-x}Sr_{x}CuO_{4}$ (LSCO) \cite{Takagi}, $Bi_{2}Sr_{1.6}La_{0.4}CuO_y$ \cite{Konstantinovic}, $YBa_{2}Cu_{3}O_{7-\delta}$ (YBCO-123) \cite{Wuyts}, and $Tl_{2}Ba_{2}CuO_{6}$ (TBCO-221) \cite{Kubo}. In all cases \emph{m} ranges from $1.1$ at $p\sim 0.2$ to $1.3$ for $p\sim 0.25$.  A fractional power law would be a clear indication of non FL behavior and has been associated with the hypothesis of a Quantum Critical Point (QCP) \cite{Naqib}, in analogy with the situation in heavy Fermions \cite{Matur}. We argue in section V that the increase of $\rho$ over linearity in overdoped samples is not necessarily related to a phase transition or a crossover, but instead may be a consequence of the physical requirement of non negative resistivity at low temperatures.

\subsection{\label{sec:level2}Hall Constant}

In contrast to the \emph{T}-independent canonical FL behavior, measurements of $R_H$ in different HTSC materials present a strong \emph{T}-dependence. $R_H$ rises as temperature goes down below room temperature, presenting a peak near $T_c$, before the normal-superconductor transition. The temperature dependence, excluding the peak itself, can be fitted to the equation $R_{H}(T) = R_{H}^{\infty} +\beta/T$ \cite{Hwang, Suzuki}, where the fit parameters depend on doping. FL theories of the normal state of the cuprates assume that the \emph{T}-dependence of $R_{H}$ decreases continuously with doping in all HTSC \cite{Pines, Levin}. However, an objective observation of published data shows on the one hand, that evolution of $R_{H}(T)$ with doping presents important differences between different HTSC systems (see discussion in section V). On the other hand, none of them fits into the canonical FL picture. The case of YBCO-123 \cite{Wuytsb}, systematically presented in this report in section III.B shows a peculiar non-monotonic behavior.

\subsection{\label{sec:level2}Hall Angle}
The Hall cotangent $cot(\theta_H) = \rho_{xx}/\rho_{xy}$ presents experimentally a robust \emph{T}-dependence, almost independent of doping, which can be fitted by $cot(\theta_H) = A + B T^2$ \cite{Wuytsb, Chien}. Some deviations from the quadratic \emph{T}-dependence have been observed, especially for doping different from optimal. Wuyts et al. \cite{Wuyts} present results for YBCO-123 films showing that $B$ increases with doping above the optimal level. This result is contrary to expected if the system were to approach the canonical FL. Konstantinovic et al. \cite{Konstantinovic} reported measurements of $\cot(\theta_H)$ in BSCCO-2212 and $Bi_{2}Sr_{1.6}La_{0.4}Cu_{1}O$, where they find the dependence $\cot(\theta_H)  = A + B T^{\gamma}$, with $\gamma$ varying with doping from $\sim 2$ for underdoped samples, going down continuously to $\sim 1.7$ in the overdoped region.

This broad spectrum of results suggests that HTSC may not approach continuously a canonical FL as doping increases and that evolution with doping varies from system to system. To our knowledge, no conclusive experimental evidence exists of a canonical FL behavior in overdoped samples, except perhaps in TlBaCuO, where $R_H(T)$ presents a weak \emph{T}-dependence and additionally the Wiedeman-Franz law has been verified \cite{Proust}. In order to explore further the anomalous overdoped state we have measured resistivity and Hall effect in $Y_{1-x}Ca_{x}Ba_{2}Cu_{3}O_{7-\delta}$  samples with different Ca and O content (doping) at temperatures above $T_{c}$. We compare our results to theoretical predictions from two representative FL models and show that they can not give a satisfactory explanation to our data.

\section{Fermi Liquid Models}

A FL phenomenological model based on the assumption of an anisotropic scattering rate along the FS was proposed by Carrington et al. \cite{Carrington}. A further development lead to the Nearly Antiferromagnetic Fermi Liquid (NAFL) of Stojkovic and Pines \cite{Pines} (hereafter referred to as SP-NAFL model). This model is supported by ARPES measurements \cite{Campuzano} and band structure calculations \cite{Andersen}, which show that the FS of hole doped HTSC looks like a square with rounded corners centered at the $\Gamma$ point ($\pi,\pi$). Two regions with different scattering rates are assumed \cite{Carrington}: hot regions corresponding to the large flat surfaces where magnetic interactions (Spin Fluctuation Scattering, SFS) are stronger, with $(\omega_h\tau_{h})^{-1}\propto T$, and cold regions near the corners of the FS, with $(\omega_c\tau_{c})^{-1}\propto T^{2}$; with $\tau_c > \tau_h$. For overdoped samples it is assumed that the FS grows with doping and that the scattering anisotropy reduces, therefore weakening the $R_H(T)$ dependence. But there are not specific quantitative predictions, in this as in most models, for overdoped samples. 

A second FL model we have chosen to compare our data to is the one from Bok and Bouvier \cite{Bouvier}, based on the Van-Hove Singularity (VHS). We will refer to it hereafter as the BB-VHS model. In this model the coexistence of electron-like and hole-like orbits at energies near the Fermi energy, when it lies close to the VHS, is emphasized. In the BB-VHS model $R_H$ is given by $R_H = \frac{1}{e} \frac{n_h(\mu_h)^2 - n_e(\mu_e)^2 }{n_h\mu_h + n_e\mu_e}$, where $n_e, n_h$ are the density of carriers and $\mu_e, \mu_h$ their mobility ($\mu_i = e\tau/m^*_i$), $i=e,h$. They have shown that including second neighbor interactions in their tight-binding calculation the FS possesses certain regions with positive and others with negative curvature. Therefore positive and negative contributions to $R_H$ may produce an apparent variable carrier density. The predicted sign change in $R_H$ would be shifted to the overdoped side due to this correction.

\section{Experimental}

\subsection{Sample preparation}
High quality c-axis oriented thin films of $Y_{1-x}Ca_{x}Ba_{2}Cu_{3}O_{7-\delta}$, with different \emph{x} and $\delta$ were prepared by DC off-axis sputtering by a method described elsewhere \cite{Krupke}. Deposition time was one and half hours, producing films with an average thickness $t\approx 1700\AA$ and an uncertainty of $\sim 13\%$. They were deposited on square $LaAlO_3$ (100) substrates of size 5x5 $mm^{2}$ or 10x10 $mm^{2}$. The doping level was adjusted by changing the Ca content \emph{x}, and/or the O content by annealing. Four Indium contacts were fixed to the corners for electrical measurements. The main source of uncertainties in our measurements come from the estimation of films thickness from the deposition time (according to previously calibrated films). This uncertainty dominates over other possible error sources, like surface roughness (typically few percents). 

\subsection{Measuring technique}
In order to preserve the square geometry of our samples (needed for later measurements of penetration length with a microwave resonator) we used the Van-der Pauw method. A current \emph{I} is sent through one diagonal and voltage is measured along the second diagonal, then we exchange the connections for \emph{I} with those for \emph{V} and measure again. This method also requires the inversion of current sense and magnetic field direction in order to compute the average. The magnetic field was applied perpendicular to the film's surface, i. e. along the c-axis, and scanned from -1 to +1 tesla. A linear fit to the data at different fields gives the slope $dV_{H}/dB$, where $B$ is the magnetic induction, at every temperature point. The Hall density of carriers is then computed as $n_{H}=[4I/et](dV_{H}/dB)^{-1}$ \footnote{What we call here the 'Hall carrier density' is computed as $n_{H}=1/eR_{H}$, where $R_H$ is measured at room temperature. We must remember that this number gives the density of carriers only for a parabolic band structure \cite{Hurd}. From $n_H$ we could try to estimate \emph{p} by computing the density of carriers per unit cell by using the 'Bond-Valence Sums' method \cite{Tallon}. However, this calculation is complicated in samples doped with Ca. An alternative method for computing \emph{p} by comparing data for $T_{c}(n_{H})$ with the universal parabola $T_{c}(p)$ is not straightforward because the relationship between $n_H$ and \emph{p} (or oxygen content \emph{x}) is not linear. A more reliable way of computing \emph{p} is by means of the universal curve for thermopower \cite{Obertelli}, but such measurements were not made here, therefore we keep $n_H$ as a representative parameter, which anyway does not affect our conclusions. Optimal doping $p_{op}$, in our samples corresponds to $n_H \approx 0.9\times 10^{22} cm^{-3}$.}, where \emph{e} is the electronic charge and $V_H$ is the Hall voltage. In the following the room temperature value of $n_H$ will be used as an indicator of the doping level. We used the Lock-in amplifier technique with an ac current of 10 mA (rms) at a frequency of 1 Hz. Resistivity is measured by sending the current \emph{I} through the contacts on one side of the sample and measuring voltage along the opposite side. A second measurement along the remaining pair of sides is necessary if the sample is not isotropic. Our samples are twinned with the a and b axis randomly distributed on the plane between the two crystallographic axes of the substrate, therefore are isotropic. Fig. 2 shows schematically the electrical connections and measurement procedure. The resistivity is then calculated as $\rho =[\pi/\ln(2)]\langle V_{R}\rangle t/I$, where $\langle V_{R}\rangle$ is the voltage averaged on the different configurations. Uncertainties on measurements of $\rho$ and $R_H$ are dominated by $t$, getting close to $15\%$. The sample is introduced in an evacuated dewar, which is immersed in a liquid nitrogen bath. Temperatures below 77K down to 60K were obtained by pumping above the cryogenic liquid. A temperature controller was used in order to stabilize the sample temperature. We have observed that a small cooling rate is very crucial in order to obtain reliable  data, in particular regarding Hall measurements vs. temperature. A cooling rate of less than 2K/minute was found to be slow enough.

\begin{figure}
\includegraphics[width=0.4\textwidth]{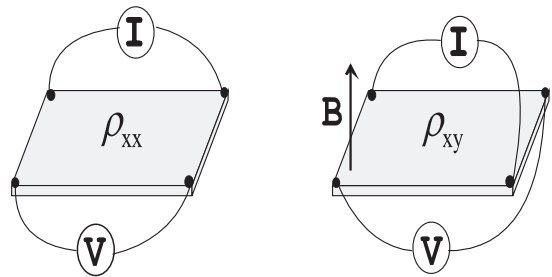}% Here is how to import EPS art
\caption{\label{fig:epsart} Schematic configuration of electrical transport measurements with the Van der Paw method. Left, longitudinal resistivity  $\rho_{xx}$. Right: transversal resistivity, or Hall constant, $R_H = \rho_{xy}/B$, where \emph{B} is the magnetic induction.}
\end{figure}

\section{Results}

\subsection{Resistivity}
Resistivity measurements of samples with different doping are presented in fig. 3. Resistivity values go down continuously with doping levels. At the highest doping level achieved, resistivity reaches 30 $\mu\Omega\cdot cm$ at 100K. At room temperature, the conductivity varies linearly with $n_H$. In order to find the temperature at which $\rho$ departs from linearity we have fitted each curve to the equation $\rho = \rho_{o} + bT$, at high temperatures. The normalized curve $[\rho(T)-\rho_o]/bT$ enhances the deviation from linearity, allowing us to better determine the temperatures $T^*$ and $T_f$. The inset in Fig. 3 presents normalized resistivity curves for some selected samples showing the departure from linearity. Resistivity of underdoped samples deviates downwards, while in overdoped samples upwards, below $T^*$ and $T_f$, respectively. Resulting values of $\rho_0$ are positive in underdoped samples, decrease with doping becoming zero at optimal doping, then become negative in overdoped samples, increasing in absolute value with doping. The slope $b$ decreases rapidly with doping in the underdoped region and remains almost constant in the overdoped regime. Fig. 4 illustrates the doping dependence of $\rho_0$ and $b$. Fig. 5 presents the data for $T^*$ and $T_f$ vs. $n_H$ obtained in the way described above. We observe an almost linear drop in $T^*$ vs $n_H$, similar to other reports \cite{Naqib}, and in agreement with the  phase diagram of Fig. 1. The line $T_f$, for overdoped samples, presents the opposite behavior, i.e. increasing with doping. 

\begin{figure}
\includegraphics[width=0.5\textwidth]{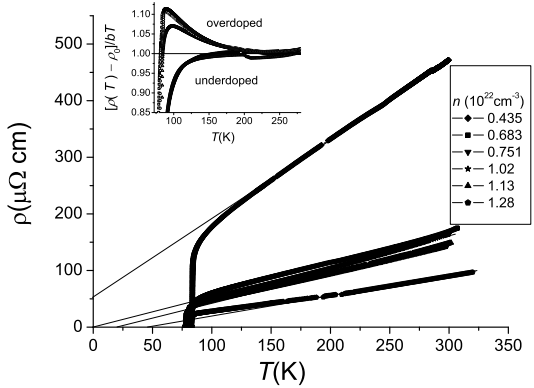}% Here is how to import EPS art
\caption{\label{fig:epsart} Measured (longitudinal) resistivity vs. temperature for samples with different doping. 
Inset: normalized resistivity emphasizing the deviation from the high temperature linear fit  $\rho(T) = \rho_o + bT$. $\rho$ deviates upwards in overdoped samples and downwards in underdoped ones.
}
\end{figure}

\begin{figure}
\includegraphics[width=0.5\textwidth]{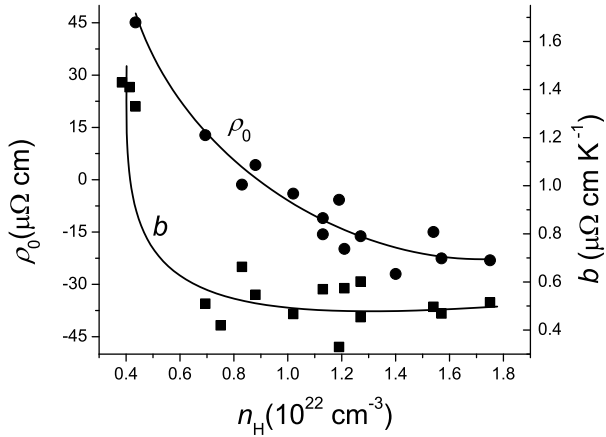}% Here is how to import EPS art
\caption{\label{fig:epsart} Fit parameters for the high temperature linear resistivity  $\rho(T) = \rho_o + bT$. Note the almost constant slope \emph{b} in the overdoped region, and  $\rho_o$ becoming zero at optimal doping.
}
\end{figure}

\begin{figure}
\includegraphics[width=0.5\textwidth]{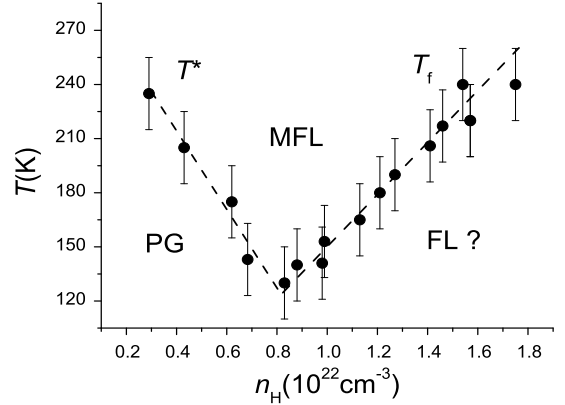}% Here is how to import EPS art
\caption{\label{fig:epsart} Characteristic temperatures for departure of $\rho(T)$ from linearity, measured from $\rho (T)$ curves in Fig. 3. The line $T*(p)$ in the underdoped region determines the pseudogap region, and $T_f(p)$ in the overdoped region determines the presumed FL phase indicated in Fig. 1.
}
\end{figure}

\subsection{Hall Effect}

\subsubsection{\label{sec:level3}Hall Constant}
Fig. 6 presents Hall Effect vs. Temperature measurements for different doping levels. The data has been fitted to the equation $R_{H}(T)=R_{H}^{\infty}[1+T_H/T]$, where $T_H$ is a characteristic temperature above which there is a weak \emph{T}-dependence. Fig. 7 presents normalized curves $R_H(T)/R_H^\infty$ vs. $T_H/T$, showing a universal temperature dependence for all doping levels. The fit parameter $T_H$ as a function of doping is shown in Fig. 8. Fig. 9 shows the results from Figs. 5 and 7, together with data for the onset critical temperature $T_{c}^{on}$, showing that $T_H$ and $T_f$ coincide. A clear conclusion is that a canonical FL can not exist in the overdoped region below the $T_H$ line, because below that temperature $R_H$ is strongly temperature dependent.

\begin{figure}
\includegraphics[width=0.5\textwidth]{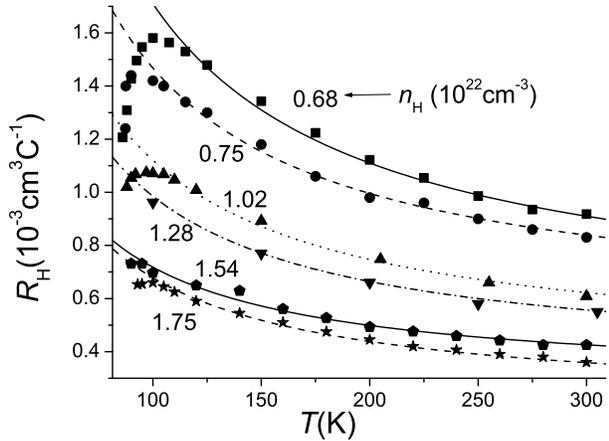}% Here is how to import EPS art
\caption{\label{fig:epsart} Measured Hall constant $R_H$  vs. temperature for samples with different doping, as indicated by the Hall number $n_H$. Note the strong \emph{T}-dependence even at the highest doping level. Curves are fits to the equation $R_H(T) = R_H^\infty (1+T_H/T)$}
\end{figure}

\begin{figure}
\includegraphics[width=0.5\textwidth]{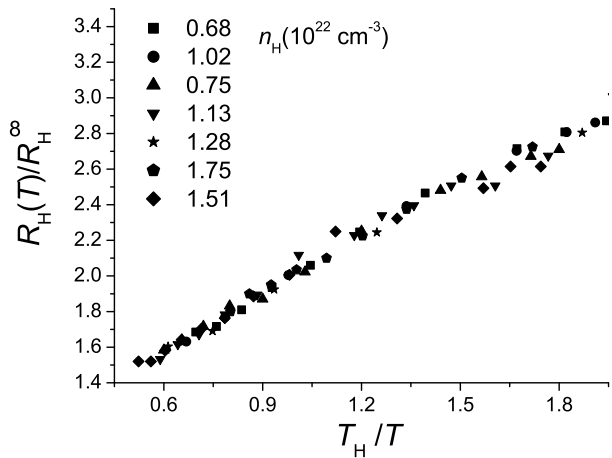}% Here is how to import EPS art
\caption{\label{fig:epsart} Measured Hall constant normalized by the fitted parameters from the equation $R_H(T) = R_H^\infty (1+T_H/T)$. We see that all the data obey quite well this law. $T_H$ is a threshold temperature above which the \emph{T}-dependency weakens. Therefore, below the $T_H$ line the system can not be a canonical FL. }
\end{figure}

\begin{figure}
\includegraphics[width=0.5\textwidth]{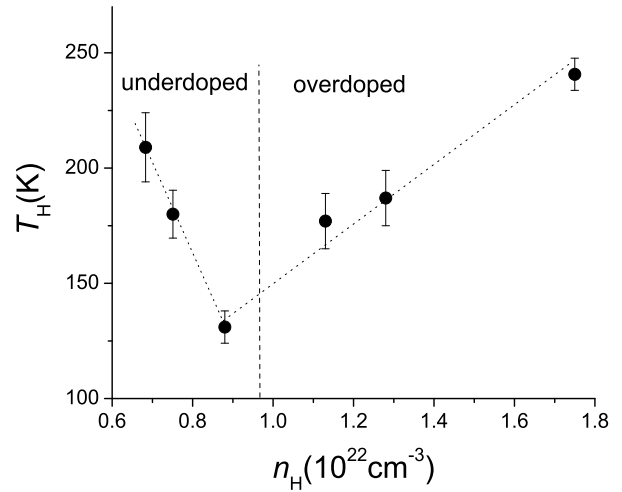}% Here is how to import EPS art
\caption{\label{fig:epsart} Fitted parameter $T_H$ vs. $n_H$. The non-monotonic, 'V' shape of this curve is reported here for the first time. This seems to be a particular behavior of the YBCO-123 HTSC system}
\end{figure}

\begin{figure}
\includegraphics[width=0.5\textwidth]{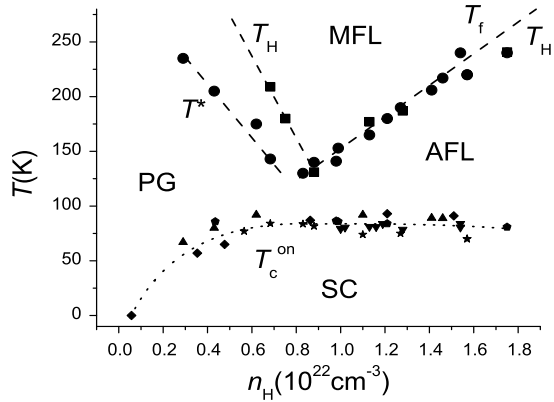}% Here is how to import EPS art
\caption{\label{fig:epsart} Phase diagram for our Ca-doped YBCO-123 samples. We have superposed the results from Figs. 5 and 8 together with measured values of the onset critical temperature $T_c^{on}$. The line $T_H$ determines the boundary between a possible canonical FL (above), and a Anomalous Fermi Liquid (AFL), below.}
\end{figure}

\subsubsection{\label{sec:level3}Hall Angle}
An important parameter we can calculate from our data is the Hall cotangent, $cot(\theta_H) = \rho_{xx}/\rho_{xy} = V_R/V_H$. Due to the cancellation of the parameters related to sample's geometry, as emphasized in the latest expression, computed values of $\cot(\theta_H)$ posses smaller relative errors, $<4\%$. Fig. 10 presents our data fitted to the quadratic law $cot(\theta_H) = A + BT^2$. A closer examination of the data reveals that this fit is not equally good for different doping levels. Fitting our data to the more general power law $cot(\theta_H) = A + BT^{\gamma}$, where $\gamma$ is now a free parameter, we obtain the dependence of $\gamma$ on doping shown in Fig. 11. 

\begin{figure}
\includegraphics[width=0.5\textwidth]{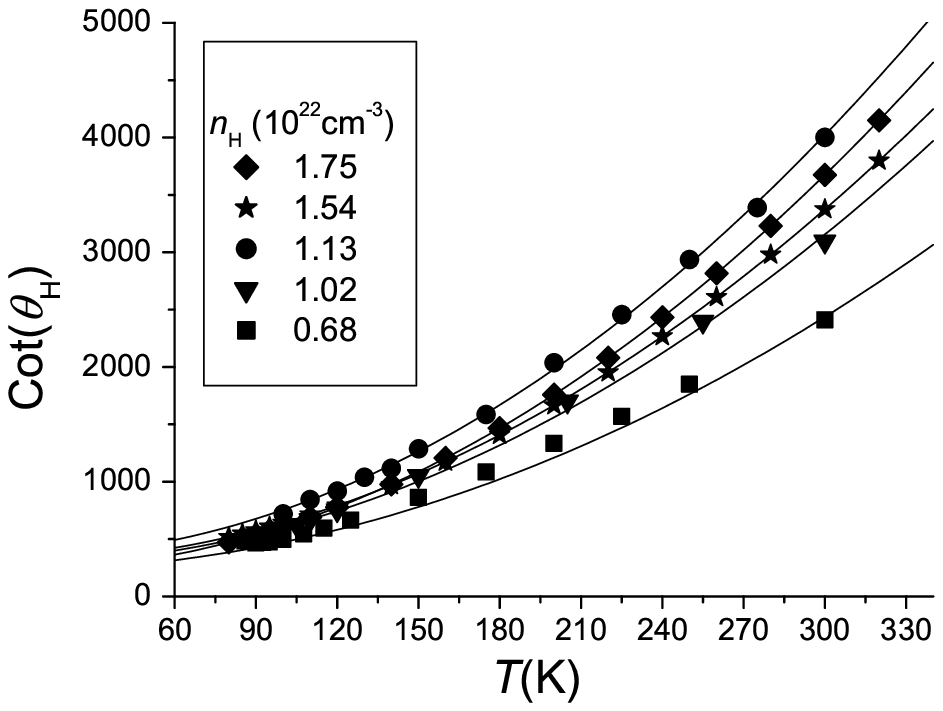}% Here is how to import EPS art
\caption{\label{fig:epsart} Hall cotangent vs. temperature for different doping, indicated by $n_H$. Curves are fittings to $cot(\theta_H) = A + BT^2$.}
\end{figure}

\begin{figure}
\includegraphics[width=0.5\textwidth]{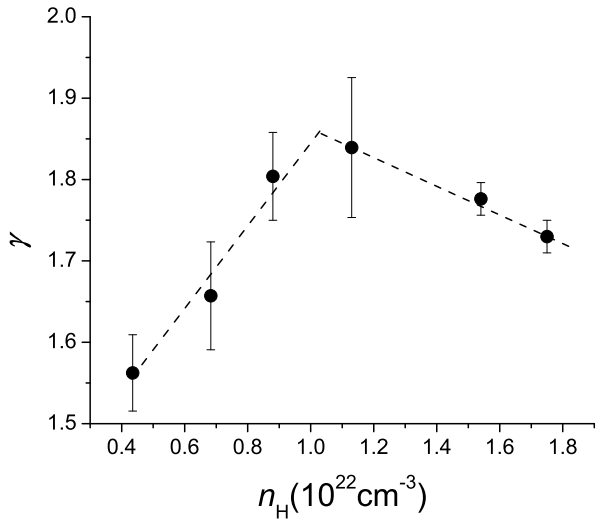}% Here is how to import EPS art
\caption{\label{fig:epsart} Parameter $\gamma$ vs. $n_H$, after fitting the data of Fig. 10 to the equation $cot(\theta_H) = A + BT^\gamma$. The exponent $\gamma$  approaches its maximum ~ 1.85 only near optimal doping.}
\end{figure}

\section{Discussion}

\subsection{Resistivity measurements}

A constant high temperature $(T > T_f)$ slope $d\rho/dT = b$ vs. doping, as we observe in the overdoped region (Fig. 4), contradicts the FL relation $\rho = m/ne^2\tau$, which implies that $d\rho/dT$ varies as $n^{-1}$. Similar behavior has been reported in \cite{Takagic} for YBCO, and for other HTSC in \cite{Levin, Suzuki}. This result reflects a unique property of the state above the $T_f$ line (MFL). It fits with the existence of a QCP \cite{Varma}, since the behavior at critical doping is recovered above the line $T_f(p)$ for $p > p_{op}$. Additional argumentation in favor of a QCP based on resistivity measurements is presented in \cite{Naqib}. The high doping region below the $T_f$ line (Fig. 1) has been studied by ARPES \cite{Kaminski}. The authors found the electronic excitations to be coherent in this region, in contrast to the non-coherent excitations in the MFL region. Such a coherence in the former region indicates a metallic (FL) behavior, as pointed out by the authors. But the introductory discussion and our data on resistivity, Hall measurements and other transport properties show an anomalous behavior not compatible with a canonical FL. Therefore, we propose to call this new phase Anomalous Fermi Liquid (AFL) (Fig.9). 

The superlinear behavior of $\rho(T)$ at $T<T_f$, sometimes adscribed to FL behavior, may have a more trivial explanation. Let us ignore for the moment the superconductive transition. Since $\rho_0$ is negative in overdoped samples, as we reduce the temperature within the linear range ($T>T_f$), $\rho(T)$ targets a negative value. Since resistivity can not be negative, at a certain temperature, which happens to be $T_f(p)$, $\rho(T)$ starts deviating upwards in order to reach a positive value at \emph{T} = 0. Given the dependence of $\rho_0$ on doping, $T_f(p)$ must also increase with doping, as observed. In this scenario $T_f(p)$ does not indicate a phase transition, nor a crossover. Hence, resistivity measurements are not conclusive regarding the FL phase. Hall effect measurements shall provide us additional criteria in order to elucidate the nature of this region.

\subsection{Hall effect measurements}

In fig. 6 we observe that all curves show a strong temperature dependence, even for the most overdoped samples.
The fitting parameter $T_H$ in Fig. 8 presents a non expected V-shape similar to that shown in Fig. 5. Above $T_H$ the system approaches the canonical FL behavior with a temperature independent $R_H$. However, this conclusion is in contradiction with our discussion of resistivity measurements. As already noted, below $T_H$ the strong temperature dependence of $R_H$ precludes a canonical FL behavior. 

The usage of unappropriate scale in reports of data like that of Fig. 6, particularly when they include very underdoped samples which possess a large peak in $R_H(T)$, may have created the misleading impression of a gradually reducing \emph{T}-dependence with doping. This impression has probably lead several researchers to conclude that, in general, overdoped HTSC are canonical FL, as pictured in the general phase diagram of Fig. 1. 
In order to show that a strong \emph{T}-dependence still remains in different overdoped HTSC systems, let us make a comparison with data published by other authors. To this effect we have computed the ratio between the peak and room temperature values, $r_H$ for different published data. For instance in TBCO-2201 the maximum \emph{T}-dependence is observed in underdoped samples where $r_{H}\approx 1.36$, and reduces gradually with doping to $r_{H}\approx 1.10$ in strongly overdoped samples \cite{Kubob, Levin}. This is in fact the HTSC system which presents the weakest \emph{T}-dependence in overdoped samples, which may indicate a smooth transition from MFL towards a canonical FL. The behavior of BSCO-2201 and $Bi_{2}Sr_{2}Ca_{1}Cu_{2}O$ (BSCCO-2212) \cite{Konstantinovic} is similar to that of TBCO with the maximum \emph{T}-dependence in underdoped samples, $r_{H}\approx 1.30$ for a hole doping $p\sim 0.05$. Then it reduces slowly to $r_{H}\approx 1.20$ in overdoped samples with $p\sim 0.23$. Therefore a substantial \emph{T}-dependence still remains at high doping. 
A contrasting behavior is found in LASCO \cite{Takagi, Levin}. At zero doping, the non-superconducting phase presents a very strong \emph{T}-dependence with $r_{H}\approx 10$, which weakens by increasing doping until it almost disappears ($r_{H}\approx 1.04$), just below the appearance of superconductivity at $p\approx 0.04$. A further increase of doping produces a reappearance of the \emph{T}-dependence, which grows proportionally to $p$. At optimal doping, $p\approx 0.15$ $r_{H}\approx 1.44$, and at high doping $p\approx 0.25$, $r_{H}\approx 1.70$. In overdoped samples the \emph{T}-dependence is more apparent at temperatures below $\sim 100 K$. Simultaneously the peak shifts to lower temperatures following the reduction of $T_c$ by doping. Above 100 K $R_{H}(T)$ looks quite flat. The case of YBCO-123 is very special. From experimental data \cite{Wuytsb} we find an already high \emph{T}-dependence, $r_{H}\approx 2.4$, at low doping, corresponding to the oxygen content $x = 6.60$. The \emph{T}-dependence is non-monotonic. There is a minimum near $x=6.85$, with $r_{H}\approx 2.14$, and an increase to $r_{H}\approx 2.5$ at maximum doping, $x\approx 7.0$. This behavior, similar to our data (Figs. 6 and 8), disagrees with the expected trend from the phase diagram of Fig. 1. 

Regarding the Hall angle, a fit to a variable exponent $\gamma$ works in fact better than the quadratic law. The fitted exponent $\gamma$ vs. $n_H$, presented in Fig. 8, attains its maximum value $\sim 1.85$, at optimal doping. It decreases faster in the underdoped side, down to 1.55 ($n_H = 0.4\times 10^{22} cm^{-3}$), and drops more slowly in the overdoped side, down to $1.73$ ($n_H = 1.7\times 10^22 cm^{3}$). This behavior can not be explained in the FL frame.

Now let us discuss our results in the light of the two FL models introduced earlier.
Let us start with the SP-NAFL model, which provides explicit expressions that allow us to evaluate temperature and doping dependences of the transport coefficients. The expression for $R_H(T)$ is a polynomial (Eq. 34 in ref. \cite{Pines}), which under the assumption $T > T_0$, where $T_0$ is a parameter in this model, can be expressed as $R_H(T) = R_H^\infty [1+6.5T_0/T]$. This is the same law as we have found, with the scaling $T_H = 6.5T_0$. $T_0$ is predicted to grow linearly with doping in underdoped samples. However, experimentally we find such a linear growth only in overdoped samples. On the contrary, in underdoped samples $T_H$ decreases with doping, as shown in Fig.8.  The 'V' shape of Fig. 8 can not be explained in this model.
The two scattering rates predicted in this model were computed for our samples and presented in Fig. 12. The magnitudes in the left scale show clearly that we are in the low field approximation, $(\omega \tau)^{-1} >> 1$. The curves follow roughly the expected temperature dependence, $(\omega_h \tau_h)^{-1}\sim T$ and $(\omega_c \tau_c)^{-1} \sim T^{2}$, with $\tau_h < \tau_c$. If $\tau$'s anisotropy were the origin for the temperature dependence of $R_H$ then the anisotropy should disappear at temperatures as low as ~ 140K for optimal doped samples (Fig. 8). The results of Figs. 8 and 9 are clearly not compatible with this assumption. We conclude that $\tau$-anisotropy alone can not explain the behavior of $R_H(T)$. A difficult point in this model is the predicted FL behavior at high temperatures. While $R_H$ satisfies that prediction for $T >T_H$, $\rho$ instead keeps its non FL linear on-temperature dependence up to high temperatures.

\begin{figure}
\includegraphics[width=0.5\textwidth]{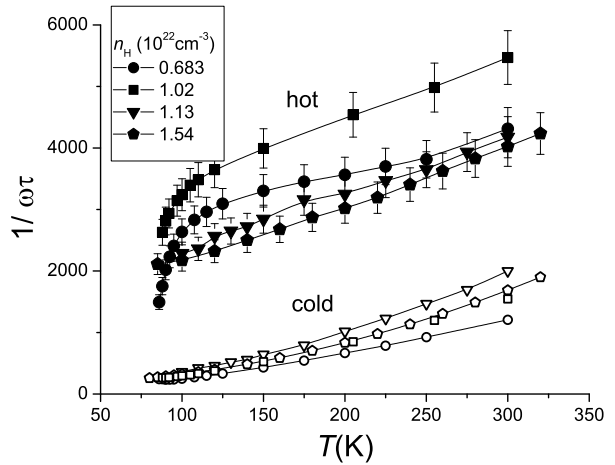}% Here is how to import EPS art
\caption{\label{fig:epsart}  Hot and cold scattering rates computed from our transport data, according to the SP-NAFL model: Cold, $1/\omega_c \tau_c = cot(\theta_H)$. Hot, $1/\omega_h \tau_h = n^2e^2\rho_{xx}R_H/B$. Lines are guides to the eye. Error bars for cold scattering rates are smaller than the symbols. The difference between both scattering rates persist at temperatures above $T_H$ (Fig. 8). This renders invalid the assumption that the \emph{T}-dependence of $R_H$ is due to $\tau$'s anisotropy (in OD samples).}
\end{figure}

Let us now turn to the BB-VHS model. Computed curves of $R_{H}(T)$, at first sight, look qualitatively similar to experiment, except for the peak, which in their calculations appears at too low temperatures, below $T_c$. This may be due to the fact that their model does not include superconductivity. However, a careful examination of their curves shows a rise of $R_H$ with reducing temperature that is much larger than experimentally observed. Computing the peak to room-temperature ratio $r_{H}\equiv R_{H}^{pk}/R_{H}(300K)$, in their plots gives $r_H \approx 1.7$ in overdoped samples, $r_H \approx 3.5$ at optimal doping, and $r_H \approx 25$ for underdoped samples. These values are above the range between 1.1 and 2.5 found for different HTSC, as reviewed above. Another difference is that their normalized curves $R_H(T)/R_H(T^*)$ vs. $T/T^*$, where $T^*$ is defined as in Fig. 5, decrease continuously, apparently to zero, at high enough temperatures. Instead, in our fit, and also as found by other authors \cite{Hwang, Wuyts}, it converges to a constant value $R_H^{\infty}$, attained already near room temperature. No results are presented for the overdoped region. A crucial point in this model is the predicted change of sign in $R_H$, related to the crossing of $E_f$ by the VHS, which should occur near $p_{op}$ \cite{Takagib}. Although experiments have confirmed a change of sign in $R_H$ at heavy doping, $p\approx 0.3$ in LASCO \cite{Takagib}, it has not been observed in our samples, nor in any other HTSC by other authors. This model predicts the correct $T^*$ vs. $p$ linear decrease. An interesting question is what would be the predicted behavior for $T_f$ or $T_H$, in the overdoped regime. The same difficulty emphasized above regarding the high temperature FL limit can not be solved in this model. In fact, we believe it will be hard in any regular FL model to harmonize the observed high temperature behaviors of $R_H$ and $\rho$.

\section{Conclusions}

Evolution of normal transport properties with doping in YBCO-123 is not monotonic, contrary to what is usually believed. Furthermore, this material, as well as other HTSC systems, does not present a clear trend towards a canonical FL at high doping. This appears to be the rule, except probably in TBCO-221. For YBCO-123 in particular we find a minimum in the \emph{T}-dependence of $R_H$ at $p_{op}$. Our measurements of the Hall effect show that $R_H(T)$ approaches a constant value, $R_H^\infty$, for any doping level at temperatures above the line $T_H(p)$. Additionally, we find no range of temperature and doping where both $\rho$ and $R_H,$ have simultaneously the canonical FL behavior. At optimal doping $R_H$ becomes rapidly \emph{T}-independent at $T > T_c$, but $\rho(T)$ is linear (MFL). In the overdoped region, at low temperatures, $R_H(T)$ has an anomalous behavior while $\rho(T)$ might be though of as having the canonical FL behavior, $\sim T^2$ (AFL). These results contradict the assumption of a canonical FL region in the overdoped region as shown in the phase diagram of Fig. 1. They can not be explained by the smooth reduction of the relaxation rate anisotropy with doping predicted by the NAFL model of Stojkovic and Pines \cite{Pines}, nor by the VHS-based model of Bock and Bouvier \cite{Bouvier}. In fact, these models, as well as others, do not address directly the problem of the overdoped regime but rather assume that it is a canonical FL. A correct model of the normal state of the HTSC should take into account the anomalous \emph{T}-dependence behavior of $\rho, R_H,$ and $cot(\theta_H)$ in the overdoped region. We encourage theoreticians to undertake the challenge of discussing these points and to extend their models to the overdoped region, which up to now has been almost neglected. 

\section {Akcnowledgements}

This work has been supported by the Heinrich Hertz center for High Temperature Superconductivity and by the Oren Family chair of Experimental Solid State Physics.


\begin{references}
\bibitem{Battloggb}  B. Batlogg and C. Varma, Phys. World Feb.(2000)33
\bibitem{Orenstein} J. Orenstein and A. Millis, Science 288 (2000) 468
\bibitem{Timusk} T. Timusk and B. Statt, Rep. Prog. Phys. 62(1999)61
\bibitem{Loram} J. Loram, K. Mirza, J. Cooper, W. Liang and J. Wade, J. Supercond. 7(1994)243
\bibitem{Varma} C. Varma and P. Littlewood, S. Schmitt-Rink, E. Abrahams and A. Ruckenstein, Phys. Rev. Lett. 63(1989)1996
\bibitem{Levin} K. Levin, J. Kim, J. Lu and Q. Si, Phys. C 175 (1991) 449
\bibitem{Tsuei} C. Tsuei, J. Kirtley, G. Hammerl, J. Mannhart, H. Raffy and Z. Li, Archiv: Cond-mat/0402655 v1 26 Feb 2004
\bibitem{Kaminski} A. Kaminski, S. Rosenkranz, H. Fretwell, Z. Li, H. Raffy, M. Randeria, M. Norman and C. Campuzano, Phys. Rev. Lett. 90 (2003) 207003
\bibitem{Friedel} J. Friedel and M. Kohmoto, Eur. Phys. J. B 30 (2002) 427
\bibitem{Battlogg}  B. Batlogg, H. Hwang, H. Takagi, R. Cava, H. Kao and J. Kwo, Phys. C, 235-240(1994)130
\bibitem{Bucher} B. Bucher, P. Steiner, J. Karpinsky, E. Kaldis and P. Wachter, Phys. Rev. Lett. 70(1993)2012
\bibitem{Ito}  T. Ito, T. Takenaka and S. Uchida, Phys. Rev. Lett. 70(1993)3995
\bibitem{Pines} B. Stojkovic and D. Pines, Phys. Reb. B (1997) 8576.
\bibitem{Proust}   C. Proust, E. Boaknin, R. Hill, L. Thaillefer and A. Mackenzic, Arxiv: cond-mat/0202101 V2 19 sept/2002
\bibitem{Naqib}  S. Naqib, J. Cooper, J. Tallon and C. Panagopoulos, Phys. C 387(2003)365
\bibitem{Wuyts}  B. Wuyts, V. Moschalkov and Y. Bruynseraede, Phys. Rev. B. 53(1996)9418
\bibitem{Chien} T. Chien, Z. Wang and N. Ong, Phys. Rev. Lett. 67 (1991) 2088
\bibitem{Takagi}  H. Takagi, B. Batlogg, H. Kao, J. Kwo, R. Cava, J. Kraweski and W. Peck, Phys. Rev. Lett. 69(1992)2975
\bibitem{Konstantinovic}  Z. Konstantinovic, Z. Li and H. Raffy, Phys. C 351(2001)163
\bibitem{Kubo} Y. Kubo, Y. Shimakawa, T. Manako and H. Igarashi, Phys. Rev. B. 43(1991)7875
\bibitem{Matur}  N. Matur, F. Grosche, S. Julian, I. Walker, D. Freye, R. Haselwimmer and G. Lonzarich, Nature 394(1998)39
\bibitem{Hwang} H. Hwang, B. Batlogg, H. Takagi, H. Kao, J. Kwo, R. Cava, J. Krajewski and W. Peck, Phys. Rev. Lett. 72(1994)2636
\bibitem{Suzuki} M. Suzuki, Phys. Rev. B, 39 (1989) 2312
\bibitem{Takagic} H. Takagi, S. Uchida, H. Uwabuchi, S. Tajima and S. Tanaka, Jpn. J. Appl. Phys. series 1 Supercond. Materials 6 (1988) 
\bibitem{Wuytsb} B. Wuyts, E. Osquiguil, M. Maenhoudt, S. Libbrecht, Z. Gao, and Y. Brunynseraede, Phys. Rev. B, 47(1993)5512
\bibitem{Carrington} A. Carrington, A. Mackenzie, C. Lin, and J. Cooper, Phys. Rev. Lett. 69 (1992) 2855
\bibitem{Campuzano} J. Campuzano, G. Jennings, M. Faiz, L. Beaulaigue, B. Veal, J. Liu, A. Paulikas, K. Vandervoort, H. Claus, R. List, A. Arko, and R. Barlett, Phys. Rev. Lett. 64 (1990) 2308  
\bibitem{Andersen} O. Andersen, O. Jepsen, A. Liechtenstein, and I. Mazin, Phys. Rev. B 49 (1994) 4145
\bibitem{Bouvier} J. Bouvier and J. Bok, Cond-mat/0203128; J. Supercond. 13 (2000) 781; Physica C 288 (1997) 217
\bibitem{Newns} D. Newns,P. Pattnaik and  C. Tsuei, Phys. Rev. B 43 (1991)3075
\bibitem{Krupke} R. Krupke and G. Deutscher, Phys. C 315(1999)99
\bibitem{Takagib} H. Takagi, T. Ido, S. Ishibashi, M. uota and S. Uchida, Phys. Rev. B. 40(1989)2254
\bibitem{Kubob} Y. Kubo, Y. Shimakawa, T. Manako, T. Satoh, S. Iijima, T. Ichihashi and H. Igarashi, Phys. C, 162-164 (1989) 991
\bibitem{Mackenzie}  A. Mackenzie, S. Julian, D. Sinclair and C. Lin, Phys. Rev. B 53(1996)5848
\bibitem{Dagan}  Y. Dagan and G. Deutscher, Phys. Rev. Lett. 87(2001)17004-1
\bibitem{Farber} E. Farber and G. Deutscher, J. Low Temp. Phys. 131(2003)563
\bibitem{Chakravarty}  S. Chakravarty, C. Nayak, S. Tewary and X. Yang, Phys. Rev. Lett. 89(2002)277003-1
\bibitem{Berhard}  C. Berhard, Ch. Niedermayer, U. Binninger, A. Hofer, J. Tallon, G. Williams, E. Ansaldo and J. Budnick, Phys. C. 226(1994)250
\bibitem{Xiao} G. Xiao, P. Xion and M. Cieplack, Phys. Rev. B.  46(1992)8687
\bibitem{Fuchs} A. Fuchs, W. Prusseit, P. Berberich and H. Kinder, Phys. Rev. B 53(1995)R14745
\bibitem{Cooper} J. Cooper, S. Obertelli, A. Carrington and J. Loram, Phys. Rev. B. 44(1991)12086
\bibitem{Markiewicz} R. Markiewicz, J. Phys. Chem. Solids 58(1997)1179 
\bibitem{Anderson} P. W. Anderson, Phys. Rev. Lett. 67 (1991) 2092.
\bibitem{Hurdb} C. Hurd, in "The Hall Effect and its Apllications", C. Chien and C. Westgate, eds. Plenum Press, 1979, p.5 
\bibitem{Kim} J. Kim, K. Levin and A. Auerbach, Phys. Rev. B 39 (1989) 11633
\bibitem{Pickett} W. Pickett, Rev. Mod. Phys. 61 (1989)749
\bibitem{Altman} E. Altman and A. Auerbach, Phys. Rev. B, 65 (2002) 104508
\bibitem{Read} N. Read and D. Newns, J. Phys. C 16 (1983) 3273
\bibitem{Coleman} P. Coleman, Theory of Heavy Fermions and Valence Fluctuations, T. Kasuya and T. Saso eds. in Springer Series in Solid State Physics vol. 62, Springer, New York 1985. 
\bibitem{Zheleznyak} A. Zheleznyak, V. Yakovenko, H. Drew and I. Mazin, Phys. Rev. B 57 (1998) 3089 
\bibitem{Ushio} H. Ushio, T. Schimizu and H. Kamimura, J. Phys. Soc. Jpn. 60 (1991) 1445
\bibitem{Fiory} A. Fiory and G. Grader, Phys. Rev. B. 38 (1988) 9198
\bibitem{Lercher} M. Lercher and J. Wheatley, Phys. Rev. B 52 (1995) R7038
\bibitem{Uemura} Y. Uemura et al. Phys. Rev. Lett. 62 (1989) 2317
\bibitem{Bernhard} C. Bernhard, C. Niedermayer, U. Binninger, A. Hofer, C. Wenger, J. Tallon, G. Williams, E. Ansaldo, J. Budnick, C. Stronach, D. Noakes and M. Blankson-Mills, Phys. Rev. B 52 (1995-II) 10488
\bibitem{Obertelli} S. Obertelli, J. Cooper and J. Tallon, Phys. Rev. B, 46 (1992) 14928
\bibitem{Tallon} J. Tallon, Phys. C 168(1990)85
\bibitem{Hurd} C. Hurd, "The Hall Effect in Metals and Alloys", Plenum Press, 1972, Chp. 4
\end{references}
     
\end{document}